\documentclass{article}
\usepackage{amsmath,epsfig}
\usepackage{algorithmic}
\usepackage{cite}
\usepackage{mathrsfs}
\usepackage{graphicx}
\usepackage{textcomp}
\usepackage{xcolor}
\usepackage{booktabs}
\usepackage{bbding}
\usepackage{color}
\usepackage{placeins}
\usepackage{float}
\usepackage[preprint]{spconfa4}

\copyrightnotice{978-1-6654-3864-3/21/\$31.00 ©2021 IEEE}

\let\OLDthebibliography\thebibliography
\renewcommand\thebibliography[1]{
  \OLDthebibliography{#1}
  \setlength{\parskip}{0pt}
  \setlength{\itemsep}{0pt plus 0.3ex}
}

\begin{document}\sloppy

\def\x{{\mathbf x}}
\def\L{{\cal L}}

\title{Lightweight Image Super-Resolution with Multi-scale Feature Interaction Network}
%
\name{Zhengxue Wang$^{\ast}$, Guangwei Gao$^{\ast}$\thanks{This work was supported in part by the National Key Research and Development Program of China under Project nos. 2018AAA0100102 and 2018AAA0100100, the National Natural Science Foundation of China under Grant nos. 61972212, 61772568 and 61833011, the Natural Science Foundation of Jiangsu Province under Grant no. BK20190089. (\textit{Corresponding author: Guangwei Gao})}, Juncheng Li$^{\dagger}$, Yi Yu$^{\ddagger}$, Huimin Lu$^{\mathsection}$}

\address{$^{\ast}$Nanjing University of Posts and Telecommunications, China; $^{\dagger}$East China Normal University, China \\
$^{\ddagger}$National Institute of Informatics, Japan; $^{\mathsection}$Kyushu Institute of Technology, Japan \\
\textit{Email: wzx\_0826@163.com, \{csggao,cvjunchengli\}@gmail.com} }


\maketitle

\begin{abstract}
Recently, the single image super-resolution (SISR) approaches with deep and complex convolutional neural network structures have achieved promising performance. However, those methods improve the performance at the cost of higher memory consumption, which is difficult to be applied for some mobile devices with limited storage and computing resources. To solve this problem, we present a lightweight multi-scale feature interaction network (MSFIN). For lightweight SISR, MSFIN expands the receptive field and adequately exploits the informative features of the low-resolution observed images from various scales and interactive connections. In addition, we design a lightweight recurrent residual channel attention block (RRCAB) so that the network can benefit from the channel attention mechanism while being sufficiently lightweight. Extensive experiments on some benchmarks have confirmed that our proposed MSFIN can achieve comparable performance against the state-of-the-arts with a more lightweight model.
\end{abstract}
\section{Introduction}
\label{sec1}

The purpose of single image super-resolution (SISR) is to attain a high-resolution (HR) image from its degraded low-resolution (LR) counterpart. For this task, many super-resolution (SR) methods have been designed and shown prominent performance, including early reconstruction-based, and more recently learning-based methods.

Recently, the solutions based on convolution neural network (CNN) have shown excellent performance in SISR task. Dong et al.~\cite{SRCNN} first presented a three-layer convolutional neural network (SRCNN) for image SR in an end-to-end manner, and achieved prominent results against previous work. Since then, many deep super-resolution networks have been proposed, such as VDSR\cite{VDSR}, MemNet\cite{MemNet}, LapSRN\cite{LapSRN}, and MSRN\cite{MSRN}. Nowadays, the attention mechanism has been widely utilized in computer vision tasks. Hu et al.\cite{SENet} proposed the squeeze-and-excitation network (SENet)  and yielded significant performance improvement in image classification task by utilizing the relationship between channels. Zhang et al.\cite{RCAN} presented the so-called residual channel attention network (RCAN) by using some residual channel attention blocks (RCAB) for image SR tasks.

However, while these complex and deep SR models can provide significant performance improvement, they can be difficult to be used to the real-world scenarios on account of the large number of parameters and computation. Especially for some mobile devices, their storage and computing resources are limited. Many works have shown that recursive networks can effectively reduce the number of parameters. For example, DRCN~\cite{DRCN} and DRRN \cite{DRRN} adopted recursive mechanism for parameter sharing. However, simply using recursion can achieve parameter reduction but at the cost of performance degradation. To handle this problem, an optional solution is to design some lightweight and effective SR models. Ahn et al.\cite{CARN} proposed a cascading residuals network for mobile scenarios (CARN-M), which reduces the computation at the cost of PSNR reduction. Hui et al.\cite{IDN} designed an information distillation network (IDN) that used a channel split strategy to aggregate current information with locally retained information. Then, on the basis of IDN, they further proposed an efficient information multi-distillation network and adaptive cropping strategy to further enhance the performance\cite{IMDN}. Lan et al.~\cite{MADNet} proposed a residual multi-scale module based on attention mechanism (MADNet) to strengthen the feature representation ability of the model. Similarly, Li et al.~\cite{sLWSR} develop a lightweight super-resolution network (s-LWSR) for efficient SISR. 

\begin{figure*}[ht]
	\centerline{\includegraphics[width=16cm]{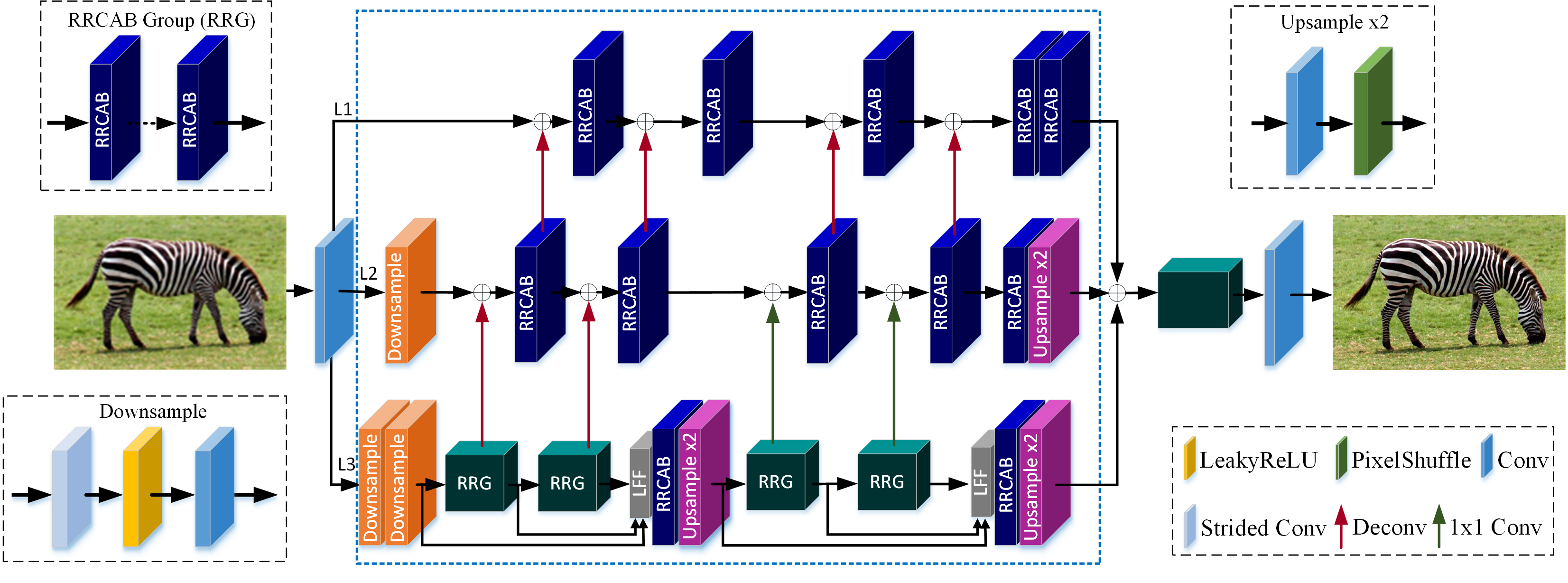}}
	\caption{The architecture of multi-scale feature interaction network (MSFIN). The dotted blue box represents the multi-scale feature interaction module (MSFIM), which consists of three levels(L1,L2 and L3), mainly composed of lightweight recurrent residual channel attention block (RRCAB) and interactive connection. $\oplus$ denotes element-wise summation.}
	\label{Figure 1}
\end{figure*}

For those aforementioned methods, there are still some issues to be addressed. First, it is found that RCAB generated a large number of parameters, which are mainly from the convolutional layers in the residual module. In addition, in multi-scale features based methods, the informative features from different scales are separately aggregated to generate the final HR features, the mutual collaboration between them are not fully exploited. Based on these observations, in order to exploit image features and restore more details, we present a  multi-scale feature interaction network (MSFIN), which can make a better trade-off between performance and the number of parameters. In summary, our major contributions can be summarized as follows: (\romannumeral 1) Based on the channel attention mechanism, we propose a lightweight recurrent channel attention block (RRCAB) for lightweight image SR task, which can improve the quality of reconstructed images with smaller memory consumption. (\romannumeral 2)  Based on these RRCABs, we develop a multi-scale feature interaction module, which can make full use of the informative features from various scales and interactive connection to better restore the details of the generated images.

\section{Proposed method}
\label{sec3}

\subsection{Network Architecture}
\label{sec31}

In this part, we describe our proposed multi-scale feature interaction network (MSFIN) in detail. As shown in Fig.1, our MSFIN is composed of three parts: the shallow feature extraction part, the multi-scale feature extraction part and the reconstruction part. In our works, the original LR image ${I_{LR}}$ is first preprocessed by the interpolation algorithm and then upsampled to the size of the target HR image. Specifically, similar to some recent super-resolution networks, a $3\times3$ convolution layer is used in the shallow feature extraction part:

\vspace{-0.2cm}
\begin{equation}
{F_{SF}} = {H_{sext}}({I_{LR}}),
\label{eq1}
\end{equation}
where ${H_{sext}}$ denotes the shallow feature extraction function and ${F_{SF}}$ are the extracted features to be sent to the multi-scale feature extraction part for deep feature learning.

The multi-scale feature extraction part is composed of deep feature extraction module and multi-scale feature interaction module (MSFIM). The multi-scale feature extraction part can be formulated as:

\vspace{-0.3cm}
\begin{equation}
{F_{DF}} = {H_{msfe}}({F_{SF}}) = {H_{dext}}({H_{msfim}}({F_{SF}}))
\label{eq2}
\end{equation}
where ${H_{msfe}}$ denotes the multi-scale feature extraction function. ${H_{dext}}$ and ${H_{msfim}}$ represent deep feature extraction function module and MSFIM respectively. ${F_{DF}}$ denotes the extracted deep features. The reconstruction part can be formulated as

\vspace{-0.2cm}
\begin{equation}
{I_{SR}} = {H_{re}}({F_{DF}}) = {H_{MSFIN}}({I_{LR}}),
\label{eq3}
\end{equation}
where ${H_{re}}$ denotes the reconstruction function, ${H_{MSFIN}}$ denote the SR network, and ${I_{SR}}$ is the reconstructed image. 

The network will be optimized with the ${L_1}$ loss function. Given a training set $\left\{ {I_{LR}^j,I_{HR}^j} \right\}_{j = 1}^M$, where $M$ denotes the number of training patches, the loss function of our MSFIN with the parameter set $\Theta$ can be formulated as

\vspace{-0.3cm}
\begin{equation}
L(\Theta ) = \frac{1}{M}\sum\limits_{i = 1}^M {||{H_{MSFIN}}(I_{LR}^j) - I_{HR}^j|{|_1}}. 
\label{eq4}
\end{equation}

\subsection{Multi-scale Feature Interaction Module}
\label{sec32}

In order to increase the receptive field of the network and extract more informative features, we devise the multi-scale feature interaction module (MSFIM). The network structure is shown in the blue dotted box in Fig.~\ref{Figure 1}, our MSFIM contains three levels: L1, L2 and L3.

\textbf{L1 level} is used to extract the shallow features with the same size as that of the HR image, and to fuse output features from corresponding stages of the L2 level. It mainly consists of several lightweight recurrent residual channel attention blocks (RRCAB). The L1 level can be formulated as

\vspace{-0.2cm}
\begin{equation}
F_{L1}^2 = H_{L1}^2(H_{L1}^1({F_{SF}} + {H_{deco}}(F_{L2}^1)) + {H_{deco}}(F_{L2}^2)),
\label{eq5}
\end{equation}
\begin{equation}
F_{L1}^4 = H_{L1}^4(H_{L1}^3(F_{L1}^2 + {H_{deco}}(F_{L2}^3)) + {H_{deco}}(F_{L2}^4)),
\label{eq6}
\end{equation}
\begin{equation}
{F_{L1}} = {H_{L1}}({F_{SF}}) = H_{L1}^5(F_{L1}^4),
\label{eq7}
\end{equation}
where $F_{Li}^j$ $(i = 1,2;j = 1,2,3,4,5)$ represents the output of each RRCAB in L1 and L2 level respectively, and $H_{L1}^n$ $(n = 1,2,3,4,5)$ denotes the RRCAB function. ${H_{deco}}$ denotes the transposed convolution (with stride 2) function. ${F_{L1}}$ and ${H_{L1}}$ represent the network output and function of the L1 level respectively.

\textbf{L2 level} is compose of a downsample block and an upsample block with a scale of 2 and several RRCABs. Different from the L1 level, we first use a $3 \times 3$ strided convolutional layer (with stride 2) followed by a LeakyReLU layer and a $3 \times 3$ convolution to conduct downsampling for input shallow features ${F_{SF}}$. Several RRCABs are then used to extract and fuse the features from L3 level. Finally, a $3 \times 3$ convolution and a sub-pixel\cite{27subpixel} convolution are applied to upsample the extracted features to the size of the HR image. The L2 level can be described as

\vspace{-0.2cm}
\begin{equation}
F_{d}^2 = {H_{d}}({F_{SF}}),
\label{eq8}
\end{equation}
\begin{equation}
F_{L2}^2 = H_{L2}^2(H_{L2}^1(F_{d}^2 + {H_{deco}}(F_{L3}^2)) + {H_{deco}}(F_{L3}^4)),
\label{eq9}
\end{equation}
\begin{equation}
F_{L2}^4 = H_{L2}^4(H_{L2}^3(F_{L2}^2 + {H_{deco}}(F_{L3}^7)) + {H_{deco}}(F_{L3}^9)),
\label{eq10}
\end{equation}
\begin{equation}
{F_{L2}} = {H_{up}}({H_{L2}}({F_{d}})) = {H_{up}}(H_{L2}^5(F_{L2}^4)),
\label{eq11}
\end{equation}
where ${H_{d}}$ and $F_{d}^2$ denote the downsample function and the downsampled features respectively. $F_{L2}^i$ $(i = 1,2,3,4,5)$ and $F_{L3}^j$ $(j = 2,4,7,9)$ denote the output of each RRCAB in L2 level and L3 level respectively. $H_{L2}^n$ $(n = 1,2,3,4,5)$ denotes the function of each RRCAB in L2 level. ${H_{up}}$ is the upsample function. ${F_{L2}}$ and ${H_{L2}}$ represent the network output and function of the L2 level respectively.

\textbf{L3 level} mainly refers to the structure of progressive upsampling\cite{LapSRN,DRN} to extract the informative features that are downsampled with a scale of 4. First of all, the extracted shallow features are operated by two downsampling blocks with a scale of 2. Then, the features are progressively fed into some RRCABs and upsampling blocks. In addition, unlike L1 and L2, we have introduced additional shortcut connection for better local feature fusion(LFF). The L3 level can be formulated as

\vspace{-0.2cm}
\begin{equation}
F_{L3}^i = H_{L3}^i(H_{L3}^{i - 1}(...(H_{L3}^1({H_{d2}}({H_{d1}}({F_{SF}}))))...)),
\label{eq12}
\end{equation}
\begin{equation}
F_{L3}^5 = H_{L3}^5({H_{LFF}}({H_{d2}}({H_{d1}}({F_{SF}}),F_{L3}^2,F_{L3}^4))),
\label{eq13}
\end{equation}
\begin{equation}
F_{L3}^i = H_{L3}^i(H_{L3}^{i - 1}(...(H_{L3}^6({H_{up1}}(F_{L3}^5)))...)),
\label{eq14}
\end{equation}
\begin{equation}
{F_{L3}} = {H_{up2}}(H_{L3}^{10}({H_{LFF}}(F_{L3}^5,F_{L3}^7,F_{L3}^8))),
\label{eq15}
\end{equation}
where $H_{L3}^i$ $(i = 1,...,10)$ represents the function of each RRCAB in L3, and $F_{L3}^j$ $(j = 1,...,9)$ represents the output of each RRCAB. ${H_{d1}}$, ${H_{d2}}$, ${H_{up1}}$ and ${H_{up2}}$ are the downsampling and upsampling functions respectively. ${H_{LFF}}$ represents the local feature fusion operation, which is composed of  concatenation operation and $1 \times 1$ convolution. ${F_{L3}}$ denotes the output of the L3 level.

Finally, the features extracted from L1, L2 and L3 levels are element-wise summed, and then sent to four RRCABs for further feature extraction. This procedure can be denoted as

\vspace{-0.2cm}
\begin{equation}
{F_{DF}} = {H_{DF}}({F_{L1}} + {F_{L2}} + {F_{L3}}),
\label{eq16}
\end{equation}
where ${H_{DF}}$ is the deep feature extraction function composed of several RRCABs and ${F_{DF}}$ is the extracted features. 

\begin{figure}[t]
	\centerline{\includegraphics[width=8cm]{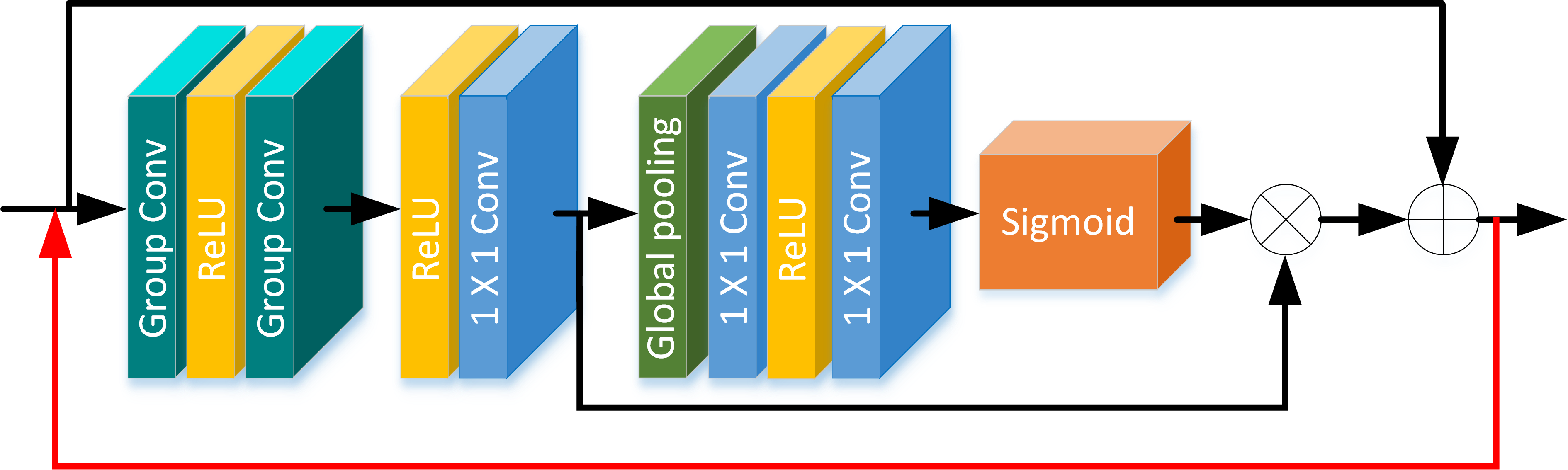}}
	\caption{The architecture of our proposed lightweight recurrent residual channel attention block (RRCAB). The red arrows indicate the loop structure. $\otimes$ denotes element-wise product.}
	\label{Figure 2}
\end{figure}

\subsection{Lightweight Recurrent RACB}
\label{sec33}

Through some experiments, it can be seen that although RCAN can achieve comparable performance, it also brings a lot of parameters and memory overhead, which is not suitable for lightweight applications. 

As shown in Fig.~\ref{Figure 2}, we design a lightweight recurrent residual channel attention block (RRCAB). In our work, grouping convolution (with group 6) is applied to replace the $3 \times 3$ convolution in RCAB, and a $1 \times 1$ convolution is added to fuse the features extracted by grouping convolution. In addition, we have added a recurrent structure to further improve the performance while remain the number of parameters unchanged. To better balance the model complexity and the quality of the generated image, each of RRCAB only looped 1 time. Based on RRCAB, we also design a more lightweight and effective SR module, named RRCAB-S, by removing the loop connections and reducing the number of the channels. The RRCAB can be described as

\vspace{-0.3cm}
\begin{equation}
{F_{ca}} = {H_{ca}}({F_{in}}) = {F_{in}} \times {H_{sig}}(H_{1 \times 1}^1(\sigma (H_{1 \times 1}^2({H_{ga}}({F_{in}}))))),
\label{eq17}
\end{equation}
\begin{equation}
{F_m} = {H_{rec}}({F_{m - 1}} + {H_{ca}}(H_{1 \times 1}^3(\sigma ({HF_{gc1}}(\sigma ({H_{gc2}}({F_{m - 1}}))))))),
\label{eq18}
\end{equation}
where ${F_{ca}}$ and ${H_{ca}}$ represent the output and function of the channel attention, respectively. ${H_{sig}}$ is the sigmoid function, and ${H_{ga}}$ denotes the global average pooling operation. $H_{1 \times 1}^i$ is $1 \times 1$ convolution, and the superscript is the index of the layers. ${H_{gc1}}$ and ${H_{gc2}}$ represent grouping convolution at different layers. ${H_{rec}}$ and $\sigma$ denote recurrent function and ReLU, respectively. ${F_{m - 1}}$ and ${F_{m}}$ represent the input and output of the module, respectively.

\section{Experiments}

\subsection{Datasets and Metrics}
\label{sec41}

As with the previous works \cite{EDSR,RCAN,MSRN,IDN}, we also train our model with the DIV2K dataset\cite{DIV2K}, which has 800 high-quality RGB training images. For testing, we also use several widely used benchmarks: Set5\cite{Set5}, Set14\cite{Set14}, BSDS100\cite{BSDS100} and Urban100\cite{Urban100}. Following the existing works, two metrics, peak signal-to-noise ratio (PSNR) and structure similarity index (SSIM)\cite{SSIM}, are utilized to validate the SR performance. We calculate the performance metrics on the luminance channel of the YCBCR color space.

\subsection{Implementation Details}
\label{sec42}

We attain the LR images by downsampling the HR images using the bicubic interpolation. To augment the training dataset, we randomly rotate and flip the 800 training images from DIV2K. In our work, the mini-batch size is set as 16, and the LR patches with a size of $48 \times 48$ are randomly cropped from the LR images as the input. We train our model with ADAM optimizer by setting ${\beta _1} = 0.9,{\beta _1} = 0.999,$ and $\varepsilon  = {10^{ - 8}}$. The learning rate is initialized to $1 \times {10^{ - 4}}$, and decreased to $6.25 \times {10^{ - 6}}$ with a cosine annealing. The proposed SR network is implemented under the PyTorch framework.

\begin{table}[t]
	\caption{Comparisons of the number of parameters and mean values of PSNR and SSIM obtained by IC, CIC and NS on four datasets $(\times 4)$.}
	\begin{center}
		\setlength\tabcolsep{5pt}
		\renewcommand{\arraystretch}{0.5}
		\begin{tabular}{@{}cccccc@{}}
			\toprule
			Scale   	&IC	 & CIC	& NS 	 &Params(K) & PSNR~~~SSIM	\\ 
			\hline
			\hline
			$\times 4$        &\XSolid   	      &\XSolid		&\XSolid         &491		   & 28.48~~~0.7951      \\
			$\times 4$   		&\Checkmark    	&\XSolid		&\XSolid        &507		   & 28.50~~~0.7961        \\
			$\times 4$        		&\XSolid    		&\Checkmark	&\XSolid        &535		   & 28.50~~~0.7962    \\
			$\times 4$     		&\Checkmark   	&\XSolid 		&\Checkmark &531	& \textbf{28.51}~~~\textbf{0.7962}   \\ \bottomrule 
		\end{tabular}
		\label{tab1}
	\end{center}
\end{table}

\begin{table}[t]
	\caption{Comparisons of the number of parameters and mean values of PSNR and SSIM with different basic modules on four datasets $(\times 4)$.}
	\begin{center}
		\setlength\tabcolsep{5pt}
		\renewcommand{\arraystretch}{0.5}
		\begin{tabular}{@{}cccccc@{}}
			\toprule
			Scale   			&CA			 &CS	              & FF	    	 	       &Params(K) & PSNR~~~SSIM	      	\\ 
			\hline
			\hline
			$\times 4$      	&\XSolid    		&\XSolid   	      &\XSolid		       &478		   & 28.46~~~0.7955          \\
			$\times 4$           &\Checkmark  	&\XSolid    		&\XSolid		       &497		   & 28.50~~~0.7959          \\
			$\times 4$      	&\Checkmark  	&\Checkmark    	&\XSolid	      		&497		   & 28.44~~~0.7943      \\
			$\times 4$     	&\Checkmark  	&\XSolid   		&\Checkmark 		&543	& \textbf{28.52}~~~\textbf{0.7967}  \\ \bottomrule 
		\end{tabular}
		\label{tab2}
	\end{center}
\end{table}

\subsection{Ablation Study}


\textbf{Comparisons of different feature interaction schemes.} On the premise that the number of channels is 20 and the size of the input LR patches is $16 \times 16$, we have carried out the following experiments, including interactive connection (IC, as shown in Fig.~\ref{Figure 1} ), complex interactive connection (CIC, add an interactive connection from Level 3 to Level 1 on the basis of IC) and interactive connection without parameter sharing (NS, deconvolution in an interactive connection does not share parameters). From Table~\ref{tab1}, we can have the following observations. First, the use of the feature interactions can improve SR performance by adding only a few parameters. Second, the CIC operation leads to an increase in the number of parameters but with little performance improvement.

\begin{table*}[htbp]
	\caption{Quantitative comparisons of MSFIN and other state-of-the-arts for scale factor 4 on Set5, Set14, BSD100 and Urban100 datasets. Best and second best results are highlighted and underlined.}
	\begin{center}
		\setlength\tabcolsep{10pt}
		\renewcommand{\arraystretch}{1}
		\begin{tabular}{@{}lcccccc@{}}
			\toprule
			 &  	&  	& Set5	 &Set14	  & BSDS100	& Urban100 \\
			Algorithm   &Scale	&Params(K)		& PSNR~~~SSIM	    & PSNR~~~SSIM 	& PSNR~~~SSIM	 & PSNR~~~SSIM	\\ \midrule
			SRCNN~\cite{SRCNN}       & 4		& 57 	& 30.48~~~0.8628        & 27.49~~~0.7503    		   & 26.90~~~0.7101    & 24.52~~~0.7221      \\
			DRRN~\cite{DRRN}      	& 4			& 297	& 31.68~~~0.8888        & 28.21~~~0.7720    		   & 27.38~~~0.7284    & 25.44~~~0.7638     \\
			IDN~\cite{IDN}     		 & 4		& 553 	& 31.82~~~0.8903        & 28.25~~~0.7730    		   & 27.41~~~0.7297    & 25.41~~~0.7632     \\
			CARN-M~\cite{CARN}  	 & 4		& 412	& 31.92~~~0.8903        & 28.42~~~0.7762   		   	   & 27.44~~~0.7304    & 25.62~~~0.7694     \\
			s-LWSR32~\cite{sLWSR}    & 4		& 571	& 32.04~~~0.8930        & 28.15~~~0.7760   		   	   & \underline{27.52}~~~\underline{0.7340}    & \underline{25.87}~~~\underline{0.7790}     \\
			MSFIN-S(Ours)  			& 4			& 352	& \underline{32.08}~~~\underline{0.8934}        & \underline{28.46}~~~\underline{0.7782}   		   	   & 27.49~~~0.7332    & 25.85~~~0.7776     \\
			MSFIN-S+(Ours)  		& 4			& 352	& \textbf{32.18}~~~\textbf{0.8945}        & \textbf{28.55}~~~\textbf{0.7798}   		   	   & \textbf{27.54}~~~\textbf{0.7344}    & \textbf{25.97}~~~\textbf{0.7804} 			    \\
			\hline
			VDSR~\cite{VDSR}   		& 4			& 665	& 31.35~~~0.8838        & 28.01~~~0.7674   		   	   & 27.29~~~0.7251    & 25.18~~~0.7524    \\
			DRCN~\cite{DRCN}       	& 4			& 1774	& 31.53~~~0.8854        & 28.02~~~0.7670   		   	   & 27.23~~~0.7233    & 25.14~~~0.7510    \\
			LapSRN~\cite{LapSRN}      & 4		& 813	& 31.54~~~0.8850        & 28.19~~~0.7720   		   	   & 27.32~~~0.7280    & 25.21~~~0.7560    \\
			MemNet~\cite{MemNet}      & 4		& 677	& 31.74~~~0.8893        & 28.26~~~0.7723   		   	   & 27.40~~~0.7281    & 25.50~~~0.7630    \\
			CARN~\cite{CARN}      	& 4			& 1592	& 32.13~~~0.8937        & \underline{28.60}~~~0.7806   & \underline{27.58}~~~0.7349    & 26.07~~~0.7837    \\
			s-LWSR64~\cite{sLWSR}    & 4		& 2277	& \underline{32.28}~~~\underline{0.8960}    & 28.34~~~0.7800		& \textbf{27.61}~~~\textbf{0.7380}    & \underline{26.19}~~~\textbf{0.7910}    \\
			MADNet~\cite{MADNet}     & 4		& 1002	& 31.95~~~0.8917        & 28.44~~~0.7780   		   	   & 27.47~~~0.7327    & 25.76~~~0.7746    \\
			IMDN~\cite{IMDN}      	 & 4		& 715	& 32.21~~~0.8948        & 28.58~~~0.7811   	           & 27.56~~~0.7353    & 26.04~~~0.7838    \\
			MSFIN(Ours)  			& 4			& 682	& \underline{32.28}~~~0.8957        & 28.57~~~\underline{0.7813}   & 27.56~~~0.7358    & 26.13~~~0.7865     \\
			MSFIN+(Ours)  			& 4			& 682	& \textbf{32.39}~~~\textbf{0.8971}      & \textbf{28.66}~~~\textbf{0.7829}   	   & \textbf{27.61}~~~\underline{0.7370}    & \textbf{26.25}~~~\underline{0.7892}     \\ \bottomrule
		\end{tabular}
		\label{tab3}
	\end{center}
\end{table*}


\textbf{Comparisons of different basic modules in RRCAB.} In this part, we mainly perform some experiments to evaluate the effectiveness of our basic RRCAB. We mainly focus on the effect of the channel attention (CA), the effect of the feature fusion (FF, $1 \times 1$ convolution followed the second group convolution, as shown in Fig.~\ref{Figure 2} ) and the channel shuffle (CS) after group convolution. FF and CS are mainly used to fuse and shuffle the features extracted by the group convolution to enhance the representation capability of extracted features. It can be seen from Table~\ref{tab2} that the CA mechanism can improve the performance with little parameters increase. The FF mechanism can improve the performance with only a few additional parameters. This indicates that our proposed RRCAB can better balance the number of parameters and the performance in a way.

\begin{figure*}[htbp]
	\centerline{\includegraphics[width=15cm]{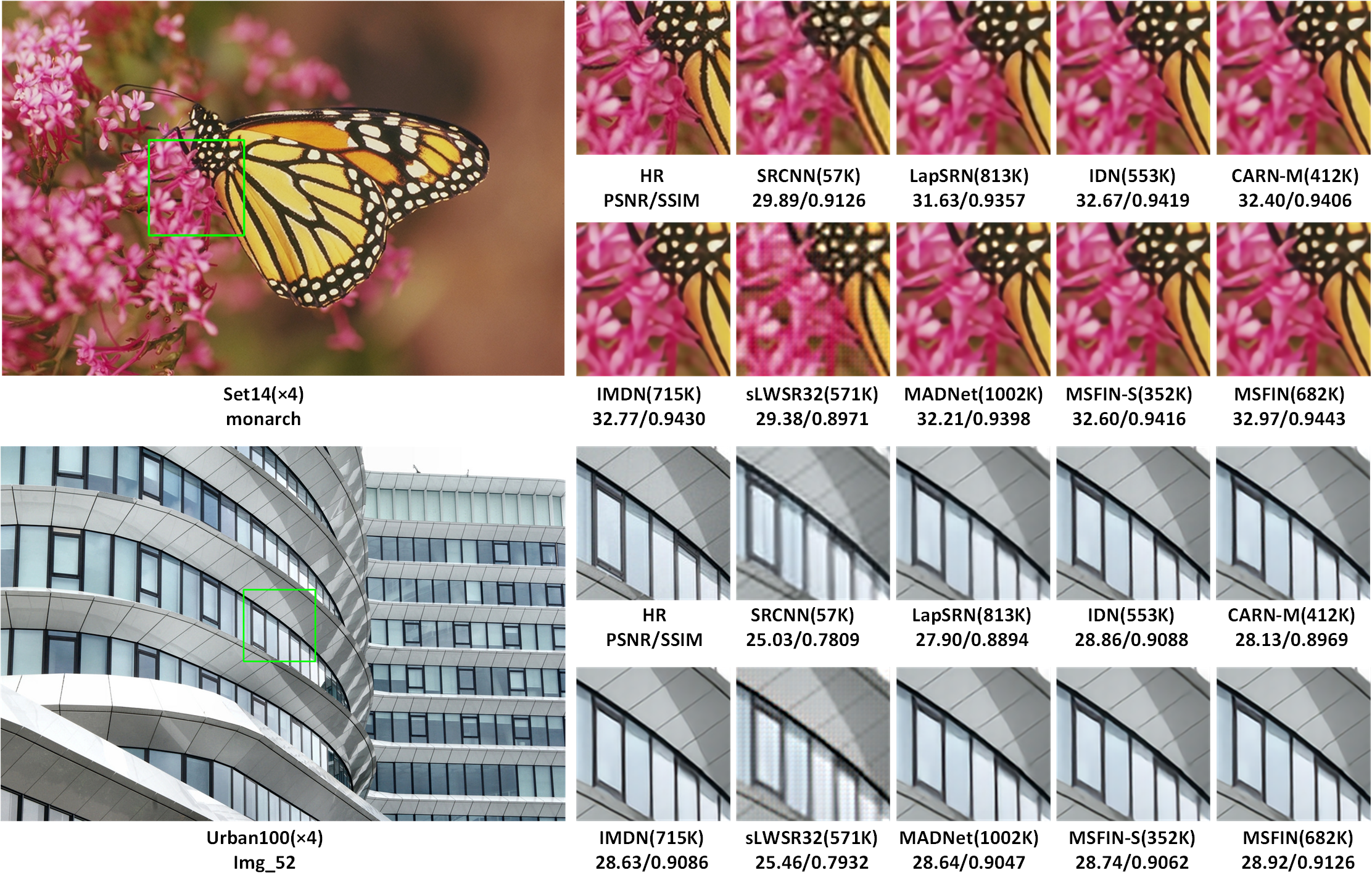}}
	\caption{Visual results of MSFIN with other SR methods on Set14 and Urban100 datasets.}
	\label{Figure 4}
\end{figure*}

\subsection{Comparison with state-of-the-arts}
\label{sec44}

To confirm the effectiveness of the proposed SR model, we compare our devised MSFIN with some state-of-the-art methods, such as SRCNN\cite{SRCNN}, VDSR\cite{VDSR}, MemNet\cite{MemNet},DRCN \cite{DRCN}, DRRN\cite{DRRN} , LapSRN\cite{LapSRN}, IDN\cite{IDN}, CARN\cite{CARN}, MADNet\cite{MADNet}, s-LWSR\cite{sLWSR} and IMDN\cite{IMDN}. In our work, in order to better assess the performance of MSFIN-S and MSFIN, we divided these methods into two groups according to the number of parameters less than 600K and greater than 600K. We mainly evaluate these methods on SR tasks with a scale of 4. As depicted in Table~\ref{tab3}, our model achieves comparable performance with acceptable number of parameters. In addition, Fig.~\ref{Figure 4} shows the visual results of each SR model on the Set14 and Urban100 datasets. It also demonstrates the effectiveness of our SR model.

In order to enhance the performance of our MSFIN, referring to RCAN, we also adopted a self-ensemble enhancement mechanism to our method, denoted as MSFIN-S+ and MSFIN+. It can be seen from Table~\ref{tab3} that our MSFIN-S+ and MSFIN+ can outperform almost all other methods within the same parameter range.

\section{Conclusions}

In this work, we devised an effective multi-scale feature interaction network (MSFIN) for lightweight single image super-resolution. In particular, we constructed a multi-scale feature interaction module (MSFIM) mainly composed of lightweight recurrent residual channel attention blocks (RRCAB), which achieved comparable performance with fewer parameters and contributed to the generation of high-quality images. In addition, MSFIM and RRCAB can expand the receptive field of the network to take full advantage of the LR image features from various scales and interactive connection. Quantitative and qualitative experiments have proved that our proposed MSFIN can have a better trade-off between performance and the model complexity.

\bibliographystyle{IEEEbib}
\bibliography{icme2021template}

\end{document}